# Ultralow thermal conductivity of single-crystalline porous silicon nanowires

Yunshan Zhao[1], Lina Yang[2], Lingyu Kong[1,3], Mui Hoon Nai [4], Dan Liu[1,3], Jing Wu[5], Yi Liu[1], Sing Yang Chiam[5], Wai Kin Chim[1], Chwee Teck Lim[4,6], Baowen Li[7,]*, John T L Thong[1,]* , Kedar Hippalgaonkar[5,]*

[1]Department of Electrical and Computer Engineering, National University of Singapore, Singapore 117583, Republic of Singapore

[2]Department of Mechanical Engineering, California Institute of Technology, Pasadena, California 91125, USA

[3]Graduate School for Integrative Sciences and Engineering, National University of Singapore, Singapore 117456, Republic of Singapore

[4]Mechanobiology Institute, National University of Singapore, Singapore 117411, Republic of Singapore

[5]Institute of Materials Research and Engineering, Agency for Science, Technology and Research, #08-03, 2 Fusinopolis Way, Singapore, 138634, Republic of Singapore

[6]Department of Biomedical Engineering, National University of Singapore, Singapore 117576, Republic of Singapore

[7]Department of Mechanical Engineering, University of Colorado, Boulder 80309, USA

∗ Corresponding authors

**Kedar Hippalgaonkar**   Tel:   +65-6501-1857
                         Email: kedarh@imre.a-star.edu.sg

**John T.L. Thong**      Tel:   +65-6516-2270
                         Email: elettl@nus.edu.sg

**Baowen Li**            Tel:   +1-303-492-0553
                         Email: Baowen.Li@colorado.edu

Keywords: Thermal conductivity, porous silicon nanowires, electron-beam technique, Young's modulus, molecular dynamics

**Abstract**

Porous materials provide a large surface-to-volume ratio, thereby providing a knob to alter fundamental properties in unprecedented ways. In thermal transport, porous nanomaterials can reduce thermal conductivity by not only enhancing phonon scattering from the boundaries of the pores and therefore decreasing the phonon mean free path, but also by reducing the phonon group velocity. Here we establish a structure-property relationship by measuring the porosity and thermal conductivity of individual electrolessly-etched single-crystalline silicon nanowires using a novel electron-beam heating technique. Such porous silicon nanowires exhibit extremely low diffusive thermal conductivity (as low as 0.33 $Wm^{-1}K^{-1}$ at 300K for 43% porosity), even lower than that of amorphous silicon. The origin of such ultralow thermal conductivity is understood as a reduction in the phonon group velocity, experimentally verified by measuring the Young's modulus, as well as the smallest structural size ever reported in crystalline Silicon (<5nm). Molecular dynamics simulations support the observation of a drastic reduction in thermal conductivity of silicon nanowires as a function of porosity. Such porous materials provide an intriguing platform to tune phonon transport, which can be useful in the design of functional materials towards electronics and nano-electromechanical systems.

**Introduction**

The urgent need for utilizing tremendous waste heat and heat management has led to extensive studies on how to tune the thermal conductivity of materials, which have wide applications in thermoelectrics, thermal sensors and other microsystems[1-7]. One way to reduce thermal conductivity is by creating ordered structures in thin films and bulk materials, called phononic crystals (PnCs), which generate a bandgap by forbidding certain frequency phonons and redistributing the phonon density of states[8-11]. An alternative approach is that of nanostructuring to reduce structural dimensions to length scales that are comparable to or less than the phonon mean free path. The thermal conductivity of bulk silicon is 150 Wm$^{-1}$K$^{-1}$ (300K), while silicon nanowires with enhanced points defects, reduced diameter, and rough surfaces can exhibit up to 100-fold reduction in thermal conductivity[3-6], which is desirable for example, if silicon is to be used for thermoelectrics[5, 7]. Porous materials can affect phonon transport by decreasing structure size and accordingly reducing the effective phonon mean free path[12]. A higher porosity with small structure size is expected to lead to lower thermal conductivity due to phonon confinement effects[13], arising from a reduced number of phonon channels as well as due to enhanced scattering of phonons at the pore interface[14].

Pore-like structures in materials could also produce non-propagating phonon modes and lower the phonon group velocity[15, 16]. Experimentally, thermal conductivity as low as 1.68 Wm$^{-1}$K$^{-1}$ at room temperature was reported by Zhang et al[17] through the fabrication of vertically aligned porous silicon arrays, albeit with unknown porosity. In addition, employing the crude kinetic theory expression[18], $\kappa = \frac{1}{3}Cvl$, where $C$ is the volumetric heat capacity, $v$ the average phonon group velocity and $l$ the phonon mean free path, the thermal conductivity scales with the phonon group velocity, which is equal to the speed of sound, $v_s$ in the low frequency limit. It has been predicted that a porosity of 30% results in a reduction in the Young's modulus, $E$ of porous silicon by half, which also leads to impeded phonon transport since $v_s \sim \sqrt{\frac{E}{\rho}}$, where $\rho$ is the mass density[19, 20]. Although simulations have predicted such phonon transport behavior in porous structures, there are still no structure-property relations that measure porosity and quantify its effects on thermal conductivity. Gravimetric measurement[21] and gas adsorption measurement[17] are useful in obtaining the average porosity across bulk porous samples, but such techniques cannot be adopted for single porous nanostructures like nanowires, nanotubes or thin films. The classical Eucken[22] and Russel models[23] are usually employed to predict thermal conductivity for bulk porous materials with periodically aligned cylindrical pores, or phononic crystals. The models assume a uniform arrangement of solid cubes and individual pores and then calculate the effect of porosity on the bulk thermal conductivity, which neglects the effect of nanoscale structure size and possible phonon-interface scattering. Therefore, not only is an accurate measurement of nanoscale porosity important, but understanding the effect of pores while accounting for structure size scattering of phonons is a key necessity.

In this work, we measure the porosity and thermal conductivity of individual silicon nanowires by employing an electron-beam technique[24]. We show that the actual cross-sectional area of the nanowire is directly proportional to the absorbed electron-beam energy. We also employ High Resolution Transmission Electron Microscopy (HRTEM) to quantify the average structure size in individual porous nanowires, and to confirm their continuously single-crystalline backbone. Then, we use an effective diffusive thermal transport model to explain the dependence of the thermal conductivity on the nanostructure size, while considering the expected change in group velocity as well as a reduced mean free path. We show experimentally that the Young's modulus of these porous silicon nanowires is smaller compared with that of bulk silicon, which translates to an expected decrease in the phonon group velocity. We also employ molecular dynamics (MD) simulations to validate the observed dependence of thermal conductivity on the porosity of these silicon nanowires. These porous silicon nanowires with ultralow thermal conductance may find new applications as TE materials.

Porous silicon nanowires were fabricated by metal assisted chemical etching (MacEtch)[25-27], as detailed in the Supporting Information S1. Figure 1a shows the as-grown porous silicon nanowires on the substrate. To make sure that the porous silicon nanowires were indeed continuously single-crystalline, Transmission Electron Microscope (TEM) diffraction pattern with a Selected Area Diffraction (SAD) aperture of 100 nm diameter was obtained and High Resolution TEM (HRTEM) characterization along the nanowire was carried out. A representative SAD pattern of porous Si nanowires is shown in the inset of Figure 1b. The HRTEM in Figure 1c shows clearly the crystalline lattice around the pores, which corresponds to the diffraction pattern. After locating such single crystalline porous Si nanowires in the TEM, the copper grid was then transferred to a Scanning Electron Microscope (SEM). Inside the SEM, a tungsten needle with a fine tip (>100 nm diameter) was used as a nano-manipulator to pick up the previously marked porous silicon nanowires from the copper grid. Once the nanowires were placed onto a micro-electro-thermal systems (METS) device, the ends of nanowires were deposited with platinum by electron beam induced deposition (EBID) to make good thermal contact, similar to the treatment used in previous work[4, 6].

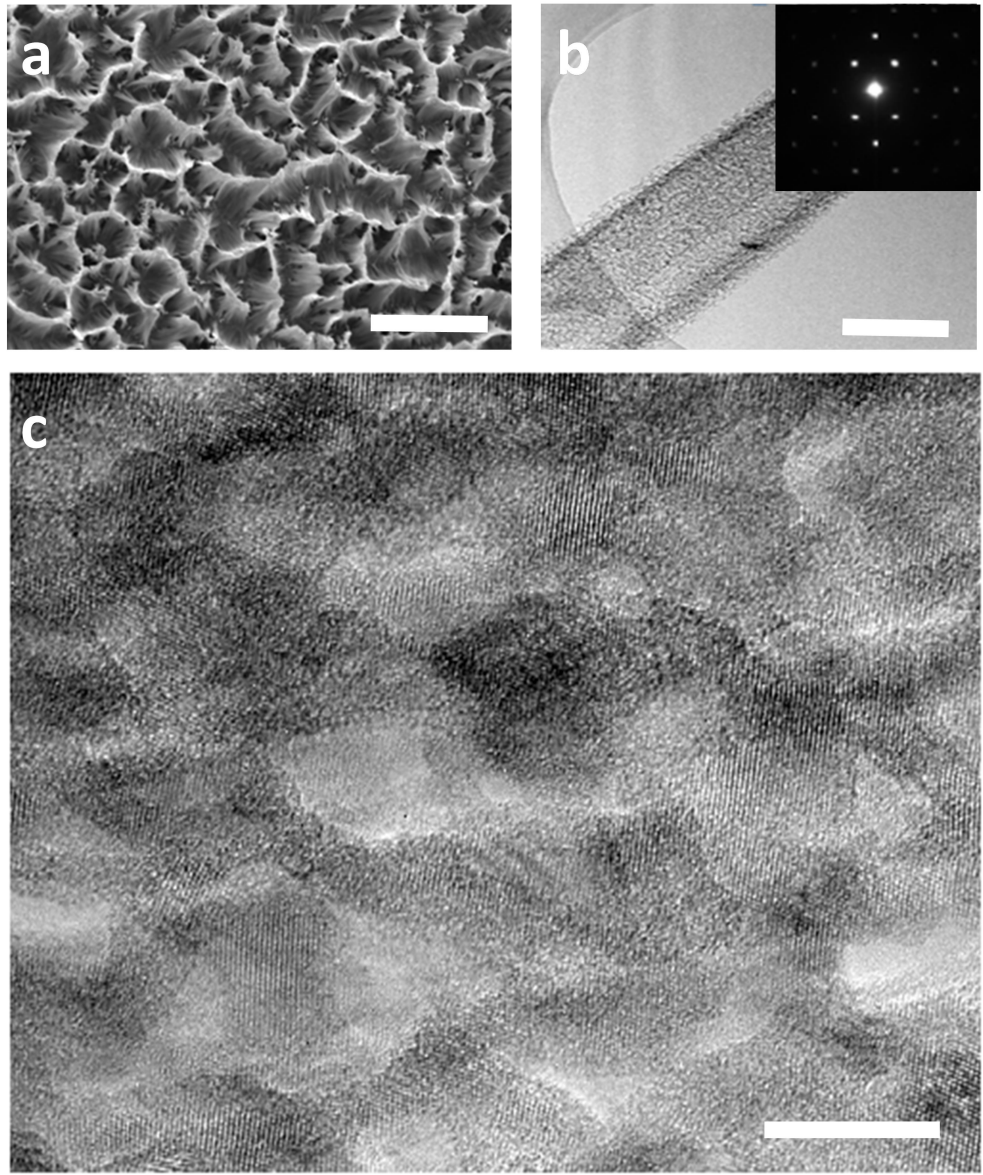

**Figure 1.** a) The as-grown silicon substrate in a top-view Scanning Electron Micrograph image. b) The porous Si nanowire on a holey carbon Transmission Electron Microscopy grid. Inset is its corresponding diffraction pattern, which shows that the nanowire is single-crystalline and the crystallinity is maintained through the length of the nanowire. c) HRTEM for porous silicon nanowire along the zone-axis b).

Scale bars are 50μm, 0.15μm and 10nm for a), b) and c), respectively.

An electron beam technique was employed to measure the thermal conductivity of single nanowires[24, 28]. Briefly, we use the focused electron beam as a heating source and the two suspended islands with platinum loops act as resistance thermometers. As shown in Figure 2a, when the electron beam irradiates the nanowire, the heat generated by the focused electron spot flows to the left and right islands, raising their temperatures to $\Delta T_L = T_L - T_0$ and $\Delta T_R = T_R - T_0$, respectively, from the original temperature, $T_0$. At thermal steady state, the heat flux from the heating spot to the islands is equal to the heat flux from the two islands to the substrate through the beams suspending the islands. It follows that the length-dependent thermal resistance, $R_i$, at any point $i$, along the suspended nanowire, is given by $R_i = R_b \left\{ \frac{a_o - a_i(x)}{1 + a_i(x)} \right\}$, where $a_o$ is the ratio of temperature rise $\Delta T_L/\Delta T_R$ when a fixed DC current is applied to the left platinum loop and the electron beam is turned off, $a_i(x)$ is $\Delta T_{Li}/\Delta T_{Ri}$ when the electron beam irradiates a spot along the nanowire at $i$. Accordingly, the thermal resistance along the porous Si nanowire length is illustrated in Figure 2b. The linearity of the fitting curve implies uniformity in the measured thermal resistance and *diffusive* phonon transport along the nanowire.

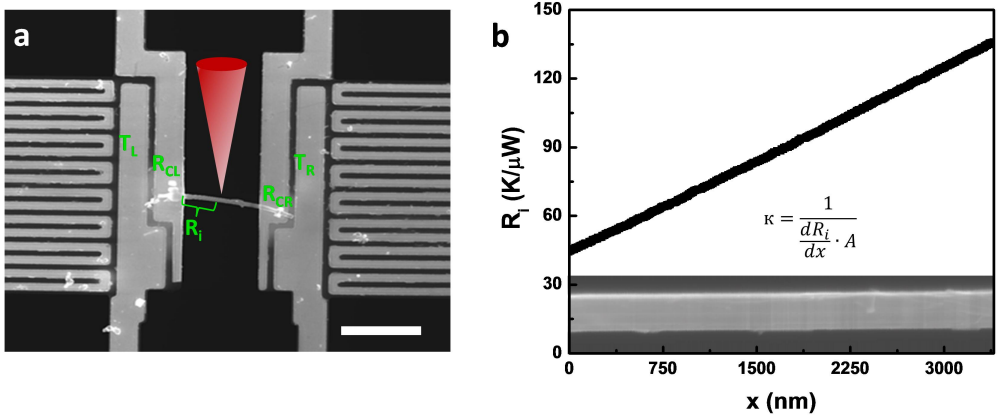

**Figure 2.** a) METS device with porous silicon nanowires in the middle. The red pyramid denotes the shining electron beam. The electron beam acts as a heat source at point *i* generating a cumulative thermal resistance in the nanowire, $R_i$, with a temperature rise of $T_L$ and $T_R$ respectively on the left and right islands. $R_{CL}$ and $R_{CR}$ are the contact resistances on the left and right contacts respectively. Scale bar for a) is 5μm. b) Cumulative thermal resistance $R_i$ as a function of distance along the nanowire. The inset below is a scanning electron micrograph of a measured nanowire. The formula shows that the thermal conductivity, $\kappa = \frac{1}{\frac{dR_i}{dx} \cdot A}$ is equal to the inverse product of this slope $\frac{dR_i}{dx}$ and the actual cross-section area, $A$.

In order to accurately determine the thermal conductivity, we need to obtain the *actual* cross-sectional area, $A$, of the porous silicon nanowire, which is defined as the solid, non-porous cross-sectional area and used for thermal conductivity calculation. The absorbed power in the e-beam technique is calculated as $E = \frac{\Delta T}{R_b}$, where $\Delta T = \Delta T_L + \Delta T_R$, is the total temperature rise measured at the islands (both left and right) and $R_b$ is the thermal resistance of all the suspending beams combined, which is a function of the incident electron beam energy and the actual cross-sectional area. Consequently, as the focused electron beam is raster scanned across the nanowire cross-section, we can deduce the dimension of an unknown specimen if its absorbed energy is measured. By using the CASINO® Monte Carlo program, we find that for a fixed electron beam energy, $E_i$, the loss of the incident electron energy scales with the cross-sectional area of the nanowires as shown in Figure 3a; the results are obtained by considering the integrated projected length across all possible electron trajectories (details in Supporting Information).

To experimentally verify the validity of this absorption power law, solid silicon nanowires with different diameters were placed on the *same* METS device as that used for the porous silicon nanowire thermal conductivity measurement. Following this, thermal contact was made using EBID to each nanowire, in a manner similar to that for the already suspended porous silicon nanowire. HRTEM imaging and diffraction were carried out as well for the solid silicon nanowires prior to manipulation (Supporting Information Figure S5). An electron beam with energy of 18keV was then scanned across each solid silicon nanowire and finally across the porous silicon nanowire. The temperature rise at both left and right islands was recorded. Four such devices with calibrated thermal resistance of the platinum beams were tested and the absorbed power was plotted as a function of cross-sectional area of the solid nanowires ($A = \pi d^2/4$, where $d$ is the diameter of the solid nanowires) as shown in Figure 3b. The linear relation between the absorption power and the cross section area of the nanowires is seen for the solid silicon nanowires on different METS devices in Figure 3b, corroborating the Casino simulation results.

In order to determine the *actual* (i.e., solid) cross-sectional area of the porous silicon nanowire, the experimentally measured absorption power is obtained from the slope, as shown by the black star in Figure 3b. The apparent cross-sectional area for the porous silicon was obtained by tilting the SEM stage by 90 degrees and imaging the cross-section, as discussed in Supplementary Information Figure S2. Therefore, porosity for the porous silicon nanowire is calculated as $P = 1 - \frac{S_{actual}}{S_{apparent}}$. The SEM image for this set-up is shown in Figure 3c and the one-to-one comparison of thermal conductivity and porosity, each measured on the same porous nanowire, is summarized in Figure 3d. Note that for the calculation of the thermal conductivity, the actual cross-sectional area determined as described above is used, not the apparent area as visible from the SEM.

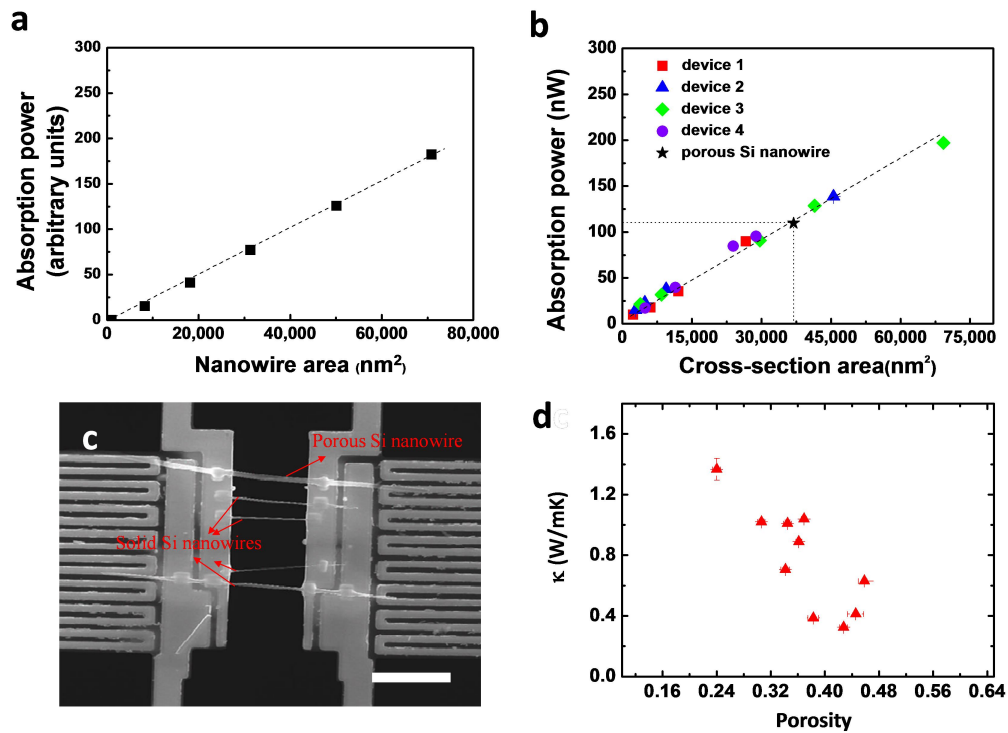

**Figure 3.** a) CASINO simulation for the power absorbed by solid silicon nanowires with different diameters. b) Experimental absorbed power plotted against the cross-sectional area of solid silicon nanowires. Also indicated is power absorption for the porous nanowire (black star): the slope of the line is used to determine the actual (solid) cross-sectional area of this wire. The energy for electron beam is 18keV. c) SEM image of the four solid silicon nanowires with different diameters and one

porous nanowire on the METS device. The scale bar is 5μm. d) Thermal conductivity as a function of porosity (each measured on the same sample) for porous silicon nanowires.

In order to understand the abnormally low thermal conductivity of porous silicon nanowires and its dependence on porosity, we employ a modified phonon radiative heat transfer (EPRT) model pioneered by Majumdar[29], where phonons are considered as energy packets moving ballistically within the crystalline part of the nanowire until scattered diffusively by the surface at the pore. In this model, the phonon mean free path is considered to be frequency-independent, called the gray-body approximation. This is important in our case of porous nanowires, since we expect that interface scattering at the pores must be the dominant scattering factor that reduces the thermal conductivity. Here a single average phonon acoustic branch (accounting for 2 transverse and 1 longitudinal branches) was considered as has been used both theoretically[30, 31] and experimentally[12] for phonon transport in porous structures with relatively small errors. This effective EPRT model provides an informative general expression for the thermal conductivity in porous materials whereby we can account for the pore morphology through the structure size, defined as $d$. The size-dependent thermal conductivity can then be written as $\kappa(T,d) = \frac{1}{3}C(T)v(\frac{1}{r_{bulk}(T)} + \frac{4}{3d})^{-1}$, where $\kappa(T,d)$ is effective thermal conductivity, $C(T)$ the volumetric heat capacity, $v$ the average phonon group velocity, $r_{bulk}(T)$ the bulk phonon mean free path and $d$ the structure size of measured samples. $C(T)$ and $v$ are taken from Chen[32], by considering each acoustic-phonon polarization and excluding all optical phonons. The factor of $\frac{3}{4}$ that multiplies the structure size, $d$, is a correction on the traditional Fourier's law, arising from solving the phonon radiative heat transfer equation and is particularly relevant since in our case as $d \ll L$, where $L$ is the length of the nanowire. The temperature-dependent bulk phonon mean free path, $r_{bulk}(T)$, is calculated by using $C(T)$ and $v$ together with the thermal conductivity of bulk silicon[33]. For porous silicon nanowires, the average crystalline size is obtained by measuring the surface periodicity of pores in HRTEM, details about which are provided in Supporting Information. Our measured structure size is around 4.3 (±1.5) nm, smaller than that reported by Gesele et al.[34] and Sui et al.[35]. Interestingly, the structure size does not seem to depend on the porosity, indicating that for larger porosity samples, the pore sizes are larger for the same cross-sectional area of the nanowire. This fit is shown as a dotted line in Figure 4a, in comparison to our data on single nanowires as well as some other experimental results on porous silicon and holey silicon thin film[8, 36-38], down to a structure size of few nanometers. It is clear that the EPRT model overestimates the thermal conductivity for all the porous silicon samples, irrespective of the sample fabrication technique. We hypothesize that for porous structures the Young's modulus is reduced considering the high surface-to-volume ratio, thus reducing the phonon group velocity[20]. To prove this, we measured the Young's modulus of our porous silicon nanowires by the commonly employed three-point bending method[39, 40] and the measurement results are shown in Figure 4c, with details summarized in Supporting Information S6. The Young's modulus of both high and low porosity silicon nanowires is a few times smaller than that of solid silicon[41, 42]. The resulting effective group velocity can then be written as $v_{eff} = \sqrt{\frac{E_{porous}}{E_{solid}}} * v$, where $E_{porous}$ and $E_{solid}$ are the measured Young's modulus for the porous and solid silicon nanowires and $v$ is the group velocity of bulk silicon. We obtain a correction to the experimental thermal conductivity due to this effective value of phonon velocity, $\kappa_{eff}(d) = \kappa_{measured}(d) * \frac{v_{eff}}{v}$. We also estimate a similar effective phonon group velocity for other porous structures shown in Figure 4a, based on $v_{eff} = (1 - P) * v$ [43, 44], where P is the reported (for literature data) porosity; the final effective thermal conductivity is shown as black open circles in Figure 4a (a summary of these results is tabulated in Table S1 in Supporting Information). This elegant result shows that over a large range of structure sizes, ranging from ~5nm (our data) up to ~500nm, the EPRT model captures the scaling of the effective thermal conductivity as a function of structure size very well and especially for our highly porous silicon nanowires. The accuracy of such a calculation can be improved further by considering frequency dependent boundary scattering of the phonons. Note here also that the bulk silicon specific heat is employed for the single crystalline porous silicon nanowires in the temperature range of our measurement, following the treatment of previous work[20], and a further discussion about the feasibility of this assumption can be found in Supporting Information S9.

To investigate the critical effect of phonon softening that manifests as a reduced group velocity we replot the measured thermal conductivity as a function of (1-P), where P is the porosity of each silicon nanowire in Figure 4b, where $v_{eff}$ is proportional to (1-P), deduced from the reduction in the Young's modulus as explained above, and the linear trend is very clearly observed. For different porosity silicon nanowires with similar structure size, the effective phonon group velocity therefore proves to be the dominant factor that tunes thermal transport. It has been shown theoretically that for nanostructured silicon samples, the group velocity of broadband phonon modes that carry heat are softened greatly as well[45] and a further discussion about the porous silicon nanowires is shown in Supporting Information S8.

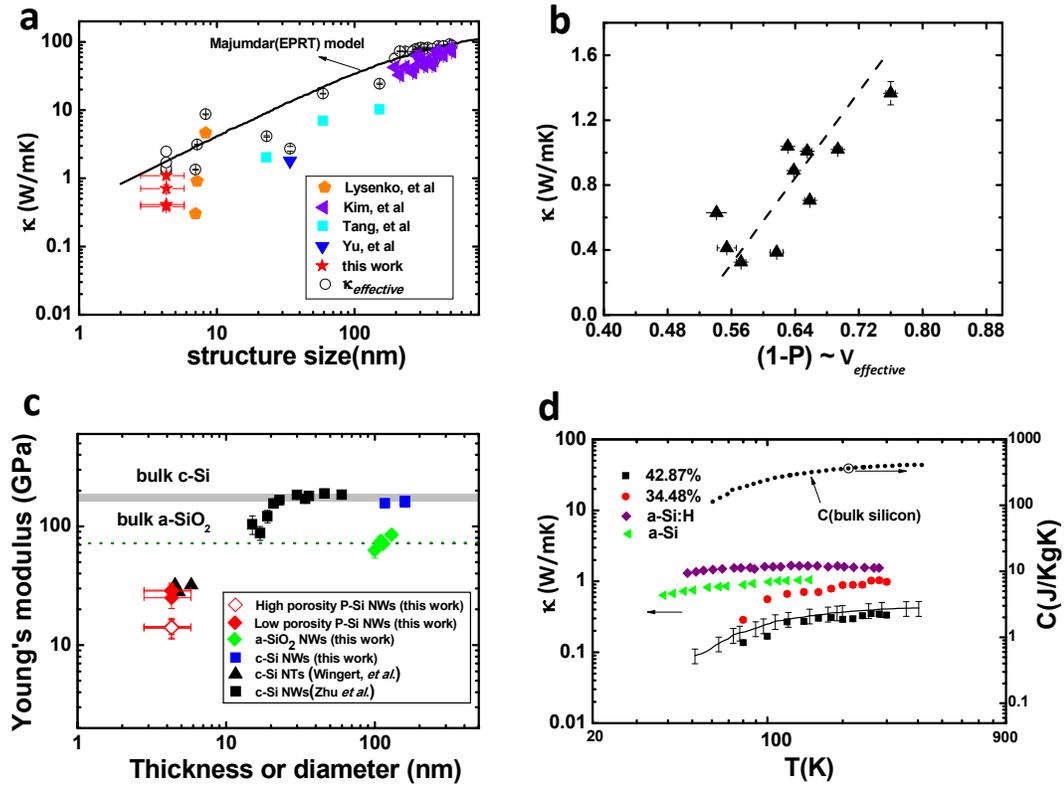

**Figure 4.** a) Thermal conductivity dependence on the structure size at 300K. The black line is the extracted effective conductivity derived from the thermal phonon radiative transport (EPRT) model from Majumdar. For comparison, holey silicon[37] (Tang et al.), porous silicon[36] (Lysenko et al.), phononic nanomesh[8, 38] (Kim et al. and Yu et al.) with different structure sizes are plotted. The effective fitting is obtained by considering porosity induced effective phonon velocity to obtain the effective thermal conductivity, $\kappa_{eff}$ and compared to the Majumdar model, as described in the text. b) Thermal conductivity plotted as a function of (1-P), where P is the porosity of the silicon nanowires. Here, we expect that the effective phonon group velocity, $v_{effective}$ is proportional to (1-P). c) The Young's modulus of our measured porous silicon nanowires, compared with crystalline silicon nanotubes (c-Si NTs)[20] and crystalline silicon nanowires (c-Si NWs)[41]. To test the measurement accuracy, we measure the Young's modulus of solid silicon nanowires with different diameters (blue solid squares) and amorphous $SiO_2$ nanowires (green solid diamonds), which is comparable to that of bulk silicon[41, 42] and bulk $SiO_2$[46]. d) Temperature-dependent thermal conductivity of porous silicon nanowire with porosity of 34.48% (red circles) and 42.78% (black squares) together with a-Si (green triangles) and a-Si:H (purple diamonds)[47]. On the right side is the temperature dependent specific heat for bulk silicon. The black solid lines are fitted from the EPRT model.

In addition, temperature-dependent thermal conductivity of two porous silicon nanowires was investigated by the traditional thermal bridge method[4, 48]. As shown in Figure 4d, the thermal conductivity of the single porous silicon nanowires is nearly insensitive to temperature in the temperature range from ~120K to 300K and decreases with temperature below ~120K, following a trend that is expected from the $C(T)$ [32], which implies that indeed the limited structural size plays the dominant role in impeding phonon transport. For these nanoporous systems, phonons with low frequency (long wavelength) have a large mean free path and are the dominant thermal carriers[49]. These phonons are expected to scatter mainly from the porous boundaries and preserve the independence of thermal conductance from temperature, as inelastic phonon-phonon scattering is expected to be weaker. This is also reflected in the EPRT model, shown as the black line[29] in Figure 4d, and suggests that the local temperature equilibrium in the nano-crystallite structure along the nanowire is maintained by thermalizing and diffusive boundary scattering. The low temperature decrease in thermal conductivity is consistent with a concomitant reduction in the specific heat.

In porous silicon nanowires with a crystalline skeleton size as small as ~5 nm, it is difficult to avoid partial oxidation on the surface of the crystalline backbone, as the native oxide layer for crystalline silicon is expected to be ~1-2 nm. Since these nanowires show ultralow thermal conductivity, a convincing verification to ensure that the nanowire is continuously crystalline is the measurement of electrical conductivity. Hence, four probe electrical conductivity measurements were conducted on the METS device. For a freshly fabricated porous silicon nanowire, electrical resistivity of 9.6 Ω•m was observed, which is five orders of magnitude larger compared to the resistivity of the silicon wafer used to etch the nanowires (~5.0x10$^{-5}$ Ω•m), similar to other reports in literature[5, 17]. In order to improve the electrical conductivity, an aluminum (Al) thin film was employed as a doping source (details of the doping are discussed in Supporting Information S4) since Al has been shown to be an effective acceptor for silicon doping[50]. The electrical resistivity of Al-doped porous silicon nanowires was measured as 5.9x10$^{-2}$ Ω•m while the thermal conductivity of the same nanowire was measured to be 0.48 Wm$^{-1}$K$^{-1}$ at 300K. Therefore, a reduction by about ~160x in the electrical resistivity with a negligible change in the thermal conductivity was observed for Al-doped porous silicon nanowires. A possible application for electrically conductive, low thermal conductivity materials is in thermoelectrics, but the electrical conductivity needs to be increased by a further 3-4 orders of magnitude in order to maximize the thermoelectric power factor, while maintaining low thermal conductivity. Post-doping[5, 51], like Boron and Gallium ion implantation, can be carried out to improve the electrical conductivity further, but is beyond the scope of our current work.

Given the large surface-to volume ratio, the morphology at the pore surface can also change the nature of scattering and hence the thermal conductivity. In order to minimize the surface energy, the free bonds at the pore surface are usually passivated by oxygen or hydrogen[52]. This passivation of atoms leads to further energy exchange between the silicon atoms by changing their vibrational spectra[47]. For porous silicon nanowires, surface oxygen passivation during electrochemical etching cannot be avoided, and can develop when exposed to atmosphere or moisture even for a few minutes. This could be a source for additional anharmonic scattering due to softening of surface phonon modes and serves as an alternative possible mechanism to explain the low thermal conductivity of the porous silicon nanowires, together with the dominant phonon-boundary scattering. Indeed, the formation of Si-O bonds at the surface of the nanowires is observed via Raman scattering, confirming the presence of such surface passivation in our porous silicon nanowires (Supporting Information S5). In order to understand better the effect of porosity on thermal transport, Nonequilibrium Molecular Dynamics (NEMD) was employed by using LAMMPS software[53, 54]. The simulation details are discussed in Supporting Information and an illustrative geometry with thermal conductivity extraction is detailed in Figure 5a. Increasing porosity leads to more phonon-interface scattering, which decreases the thermal conductivity sharply. Figure 5b shows the thermal conductivity dependence on the porosity for porous silicon nanowires and the inset is a silicon nanowire with porosity of 43.6%. The normalized thermal conductivity (with respect to the thermal conductivity of a crystalline silicon nanowire of the same diameter) as a function of porosity is shown in the inset of Figure 5b. The thermal conductivity of the porous silicon nanowire with a porosity of 43% is decreased by one order of magnitude compared to that of a solid silicon nanowire due to the scattering by pores. A reduction in thermal conductivity by a factor of ~3 is observed when the porosity is increased from 22% to 43%, which is in line with our experimental observation at 300K, with the thermal conductivity reduced by a factor of ~4 for a porosity increase from 25% to 40%. Similar to the experimental observation of a reduced Young's modulus in porous silicon nanowires, we expect a softening effect in the simulations too[55]. The Young's modulus of the same porous silicon structure can be obtained by NEMD with the simulation details shown in Supporting Information Figure S11 and S7. Here, the Young's modulus is shown to decrease strongly as the porosity increases, leading to a concomitant decrease in the phonon group velocity. In correspondence with the experimental conditions, the oxygen passivation effect is investigated as well, which decreases the Young's modulus further for the same porosity. Note here that a direct comparison with (1-P) similar to the experiments is not possible since the pore distribution in the NEMD structures are not homogeneous, unlike in our porous nanowires.

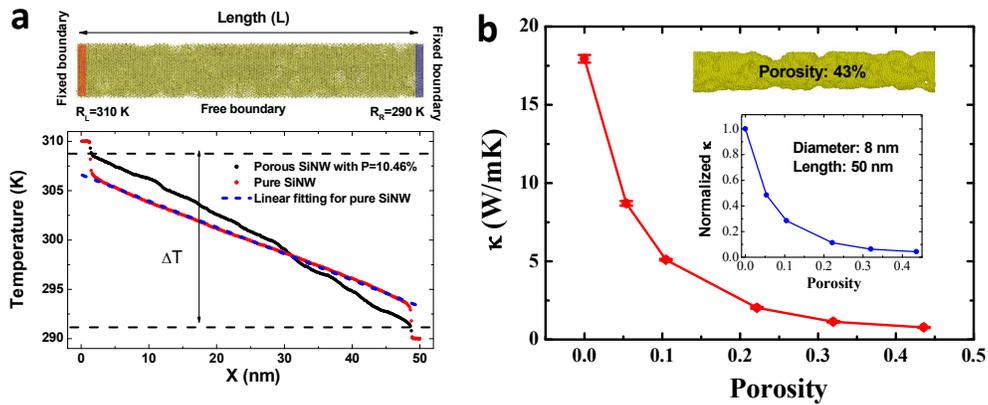

**Figure 5.** a) Structure of porous silicon nanowire with porosity 10.46% is shown in the upper panel, the corresponding temperature profile of this nanowire is shown in lower panel. Free boundary condition is used for the surface of silicon nanowire and fixed boundary condition is used for direction along silicon nanowire. Two heat baths are set at the ends of silicon nanowire with high temperature (310 K) and low temperature (290 K), respectively. The diameter is 8nm and length is 50nm. b) Thermal conductivity of porous silicon nanowire versus porosity. Inset is the structure of porous silicon nanowire with porosity of 43% after relaxation. The normalized thermal conductivity, which is defined as the ratio of thermal conductivity of a porous silicon nanowire to that of solid silicon nanowire, is shown in the inset. The value at zero porosity corresponds to the thermal conductivity of solid silicon nanowire. The thermal conductivity decrease rapidly as the porosity increases to 22.15%, while decreases slowly as the porosity increases from 22.15% to 43.59%.

In conclusion, single-crystalline porous silicon nanowires have been fabricated to study the effect of porosity on thermal transport. By means of electron-beam based power absorption measurements on single nanowires, their porosity is experimentally obtained and the thermal conductivity is measured on the same nanowires. The small structure size (~5nm) as well as the softening effect due to reduced Young's modulus of porous silicon nanowires impedes phonon diffusive transport, giving a thermal conductivity as low as 0.33 Wm$^{-1}$K$^{-1}$ at 300K, while maintaining modest electrical conductivity by Al-doping. Our study shows that by enhancing phonon-surface scattering in nanostructures with a large surface-to-volume ratio, it is possible to tune thermal conductivity in single-crystalline materials below their amorphous analogs, which is not only interesting fundamentally, but can also be very useful in applications such as insulation barriers for thermal management applications and in thermoelectrics. Mapping internal structure of such homogeneously porous nanowires to thermal conductivity will be of both theoretical and experimental interest for future studies.

**Supporting Information**

Supporting Information is available from the Wiley Online Library or from the author.